\documentclass[prl,aps,twocolumn,superscriptaddress,amsmath,amssymb, showpacs]{revtex4}
\pdfoutput=1

\usepackage{epsfig}
\usepackage{epsfig}	

\newcommand{\ket}[1]{| #1\rangle}

\newcommand{\beqa}{\begin{eqnarray}}
\newcommand{\eeqa}{\end{eqnarray}}
\newcommand{\beq}{\begin{equation}}
\newcommand{\eeq}{\end{equation}}

\begin{document}

\title{Ancilla-less selective and efficient quantum process tomography}

\author{Christian Tom\'as \surname{Schmiegelow}}
\affiliation{Departamento de F\'{\i}sica \& IFIBA, 
FCEyN, UBA, Pabell\'on 1, Ciudad Universitaria, 1428 Buenos Aires,
 Argentina}
 
\author{Ariel \surname{Bendersky}}
\affiliation{Departamento de F\'{\i}sica \& IFIBA, 
FCEyN, UBA, Pabell\'on 1, Ciudad Universitaria, 1428 Buenos Aires,
 Argentina}

\author{Miguel Antonio \surname{Larotonda}}
\affiliation{CEILAP, CITEDEF, J.B. de La Salle 4397, 1603 Villa Martelli,
Buenos Aires, Argentina}

\author{Juan Pablo \surname{Paz}}
\affiliation{Departamento de F\'{\i}sica \& IFIBA, 
FCEyN, UBA, Pabell\'on 1, Ciudad Universitaria, 1428 Buenos Aires,
 Argentina}

\begin{abstract}
Several methods, known as Quantum Process Tomography, are available to characterize the evolution of quantum systems, a task of crucial importance. However, their complexity dramatically increases with the size of the system. Here we present the theory describing a new type of method for quantum process tomography. We describe an algorithm that can be used to selectively estimate any parameter characterizing a quantum process. Unlike any of its predecessors this new quantum tomographer combines two main virtues: it requires investing a number of physical resources scaling polynomially  with the number of qubits and at the same time it does not require any ancillary resources. We  present the results of the first photonic implementation of this quantum device, characterizing quantum processes affecting two qubits encoded in heralded single photons.  Even for this small system our method displays clear advantages over the other existing ones.
\end{abstract}

\maketitle

\paragraph*{Introduction.}

Quantum process tomography (QPT) \cite{NC, emerson2007symmetrized, BPP08, BPP09,SLP10, mohseni2007direct, mohseni2006direct, cecilopez1, cecilopez2,LBPC10} is a task of fundamental and practical importance. In fact, it can be used not only to study the evolution of generic quantum systems but also provides the necessary information to  characterize the most important noise sources affecting a quantum information processor. This task is essential to design good error correction strategies which must be used in order to prevent the decoherence induced by the coupling with the environment. 
 
The full characterization of a quantum process (i.e., full QPT) is always a hard task that requires investing resources that scale exponentially with the number of qubits (such number is hereafter denoted as $n$, being $D=2^n$ the dimension of the corresponding set of states). As the complete characterization of a quantum process requires $O(D^4)$ real numbers (see below), it is necessary to have methods that enable a partial characterization of quantum processes in an efficient manner (i.e., investing resources scaling polynomially with the number of qubits of the system). Ideally, two requirements must be imposed on a good method for QPT: First, the method should enable us to select only a few parameters of the process to be efficiently determined. Also, the method should avoid the use of expensive resources such as clean qubits (i.e., qubits that are required to be perfectly isolated from the environment even though they interact with the system). Existing methods for QPT do not satisfy these criteria: They are either inefficient or they use expensive ancillary qubits. In this report we present the first method satisfying the two requirements. In fact, our method can be used to perform partial QPT without using any ancillary qubits while efficiently estimating any parameter of the $\chi$ matrix of a quantum process (with accuracy independent on the number of qubits of the system).

In essence the method maps any relevant parameter of a quantum process onto the average transition probabilities between a special set of quantum states through a quantum process. Such states can be efficiently prepared, sampled and detected. Using a heralded single photon source we fully implement this method to perform QPT on any process jointly affecting the polarization and momentum qubits of single photons.

\begin{figure}[t]
 \begin{center}
\epsfig{file=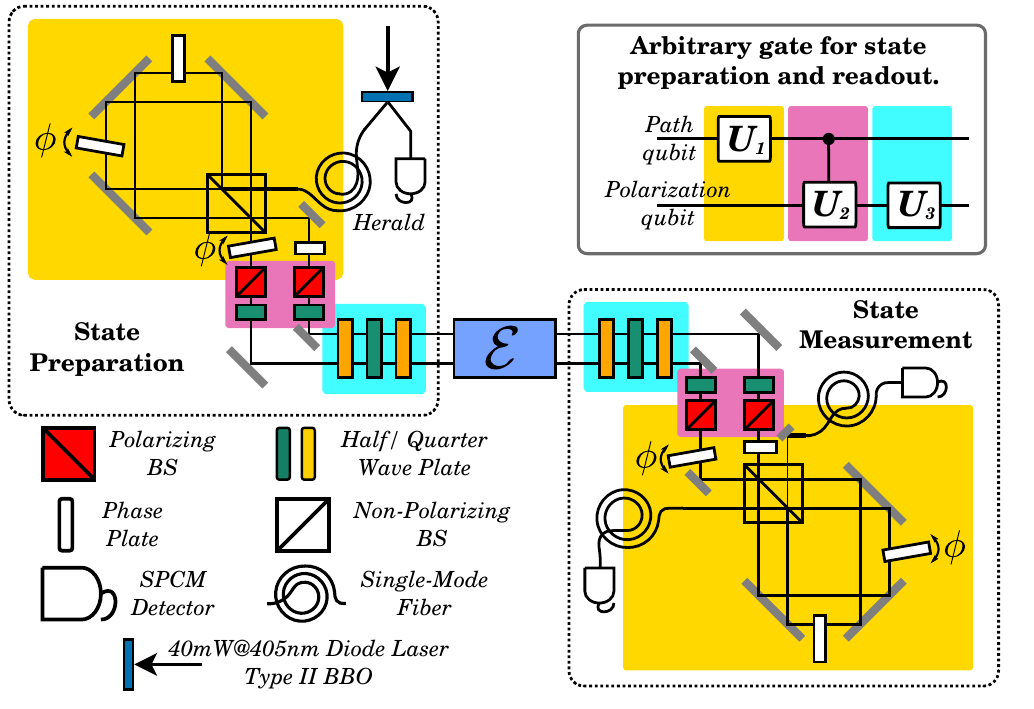, angle=0, width=0.450\textwidth}
\end{center}
\caption{\textbf{Experimental setup.} Single $810nm$ heralded photons are generated at a BBO cristal by Type II parametric down conversion. Arbitrary states are prepared and measured with a combination of single qubit unitary gates and a controlled operation as exemplified in the inset, colours show the correspondence between circuit and physical implementation. Path-qubit unitary gates are controlled with phase plates and Sagnac interferometers. Polarization-qubit unitary gates are achieved with a combination of wave plates affecting both paths. A controlled operation with control on path and target in polarization is implemented with wave plates at  different angles in each path. The different processes studied were set up in the zone marked as $\mathcal{E}$. Single-Mode fibers clean the photon's spatial mode ensuring good interferometer visibility ($>95\%$ in the Sagnacs and $\approx90\%$ in the Mach-Zehnder). Not shown in the figure is the Mach-Zehnder's active phase-sensitive stabilization mechanism (see supplementary material).}
\label{fig:scheme}
\end{figure}

\paragraph*{Ancilla-less quantum process tomography.}

An unknown quantum process affecting a physical system composed of $n$ qubits can be represented as a linear map (a quantum channel) taking initial states $\rho_0$ into final states $\rho=\Lambda(\rho_0)$. Any such map can be parametrized in terms of a $\chi$--matrix as follows: Given a basis of operators $E_a$ (with $a=0,...,D^2-1$) any channel can be written as  $\Lambda(\rho)=\sum_{a,b}\chi_{ab}E_a\rho E^\dagger_b$. The $D^2\times D^2$ matrix $\chi$ parametrizes the map which. It is hermitian and trace preserving if and only if  $\chi=\chi^\dagger$ and $\sum_{ab}\chi_{ab}E^\dagger_b E_a=I$. Positivity of the $\chi$--matrix is equivalent to the complete positivity (CP) of the map $\Lambda$. Simple counting arguments show that the number of real parameters defining such map is $O(D^4)$. This makes the task of full QPT hopeless for large systems (unattainable even for systems of a few tens of qubits). Moreover, until recently, methods available for QPT \cite{NC} were such that the evaluation of a sub-set of elements $\chi_{ab}$ also required resources scaling exponentially with $n$. In fact, the standard approach to QPT consists of estimating the transition probabilities $P_{kk'}=Tr(\rho_{k'}\Lambda(\rho_k))$ for a complete set of initial and final states $\rho_k$ and $\rho_{k'}$ (with $k, k'=0,...,D^2-1$). After this exponentially large number of experiments the $\chi_{ab}$ elements are determined by inverting a set of linear equations which is also exponentially large. Improvements in this inefficient method were presented recently \cite{BPP08,BPP09,SLP10}. Such methods enable the efficient estimation of any diagonal element $\chi_{aa}$. However, to estimate off-diagonal elements $\chi_{ab}$ such methods require the use of extra ancillary clean qubits. We now present a general method that overcomes this problem. It is efficient and it does not require the use of extra clean qubits. The method is based on an important fact: any element of the $\chi$--matrix can be interpreted as the average survival probability (i.e. fidelity) of a certain quantum map \cite{BPP09}:
\beq
F_{ab}\equiv  \int d|\phi\rangle \ \langle\phi| \Lambda(E_a^\dagger |\phi\rangle\langle\phi| E_b) |\phi\rangle={{D\chi_{ab}+\delta_{a,b}}\over{(D+1)}}.\label{ecuacion1}
\eeq 
This is the average over the entire Hilbert space of the survival probability of a map $\Lambda_{ab}$ defined as $\Lambda_{ab}(\rho)=\Lambda(E_a^\dagger\rho E_b)$ (i.e., it is obtained by first transforming $\rho$ into $E_a^\dagger\rho E_b$ and then applying the channel $\Lambda$). The efficient estimation of $F_{ab}$ is equivalent to that of $\chi_{ab}$. However, two main obstacles are apparent impediments for the efficient estimation of $F_{ab}$. The first is that averaging over the entire Hilbert space apparently requires preparing and measuring a infinite number of quantum states. The second obstacle is that the effective channel $\Lambda_{ab}$ is not physical (it is generally not a CP map unless $E_a=E_b$). The first obstacle can be surmounted by using the tools presented in \cite{BPP08,BPP09}. Thus, we can transform the integral over the entire Hilbert space into a sum over a finite set of states that form a so-called 2-design, which exist for any dimension \cite{2design,renes2004,Dan05:MT,mcconnell2007efficient,DCEL06,klappenecker2005mutually}. In fact, if the set $S=\{|\phi_j\rangle, j=1,...,K\}$ is a $2$--design 
\beq
F_{ab}={1\over K} \sum_j \langle\phi_j| \Lambda(E_a^\dagger |\phi_j\rangle\langle\phi_j| E_b) |\phi_j\rangle.
\eeq 
The exact computation of $\chi_{ab}$ involves finite but exponentially large resources since $K=O(D^2)$. However, by randomly sampling over a subset of the $2$-design, after $M$ experiments one estimates $F_{ab}$ with an error that scales as $\Delta F_{ab}\propto\sqrt{\frac{1}{M}\left(1-\frac{M-1}{K-1}\right)}$. The error scales roughly as $1/\sqrt{M}$ for $M\ll K$ and vanishes for $M=K$ \cite{BPP08,BPP09,SLP10}. Thus, the precision fixes the required number of experiments,  not the size of the Hilbert space. 

The way to surpass the second obstacle, i.e. the fact that $\Lambda_{ab}$ is not a physical map, is to notice that it can be obtained as the difference between two CP maps. To be precise, let us describe how to achieve the efficient estimation of the real part of $\chi_{ab}$. We exploit the connection with the real part of $F_{ab}$ shown in (2) and define the fidelities $F_{ab}^\pm$ of two efficiently obtainable CP maps (see below) as: 
\beq
F_{ab}^\pm= \sum_j \langle\phi_j| \Lambda((E_a\pm E_b)^\dagger |\phi_j\rangle\langle\phi_j| (E_a\pm E_b)) |\phi_j\rangle. 
\eeq
The desired fidelity is obtained by measuring $F_{ab}^\pm$ and using that $2Re(F_{ab})=F_{ab}^+-F_{ab}^-$. Therefore the estimation of $F_{ab}$ (and with it, the estimation of $\chi_{ab}$) is summarized as follows: 1) Randomly choose an element of the $2$--design $|\phi_j\rangle$; 2) Efficiently prepare the state obtained by acting with $(E_a\pm E_b)^\dagger$ on the state chosen in the first step (see below); 3) Apply the channel $\Lambda$ to the resulting state; 4) Estimate the probability to detect $|\phi_j\rangle$ as the output state. By repeating this process $M$ times we estimate $F_{ab}^\pm$ with an accuracy scaling as $1/\sqrt{M}$. This is the core of the method that is the first one enabling the efficient estimation of any $\chi$--matrix element without using extra ancillary resources. 
The above steps are applicable because: i) the states of the $2$--design, $|\phi_j\rangle$, can be efficiently prepared and detected; and ii) the states obtained by acting with $(E_a\pm E_b)^\dagger$ on $|\phi_j\rangle$ can be prepared. The details of this preparation process and other technical remarks including determination of the imaginary $\chi$ elements can be found in the supplementary material.

\begin{figure*}
\epsfig{file=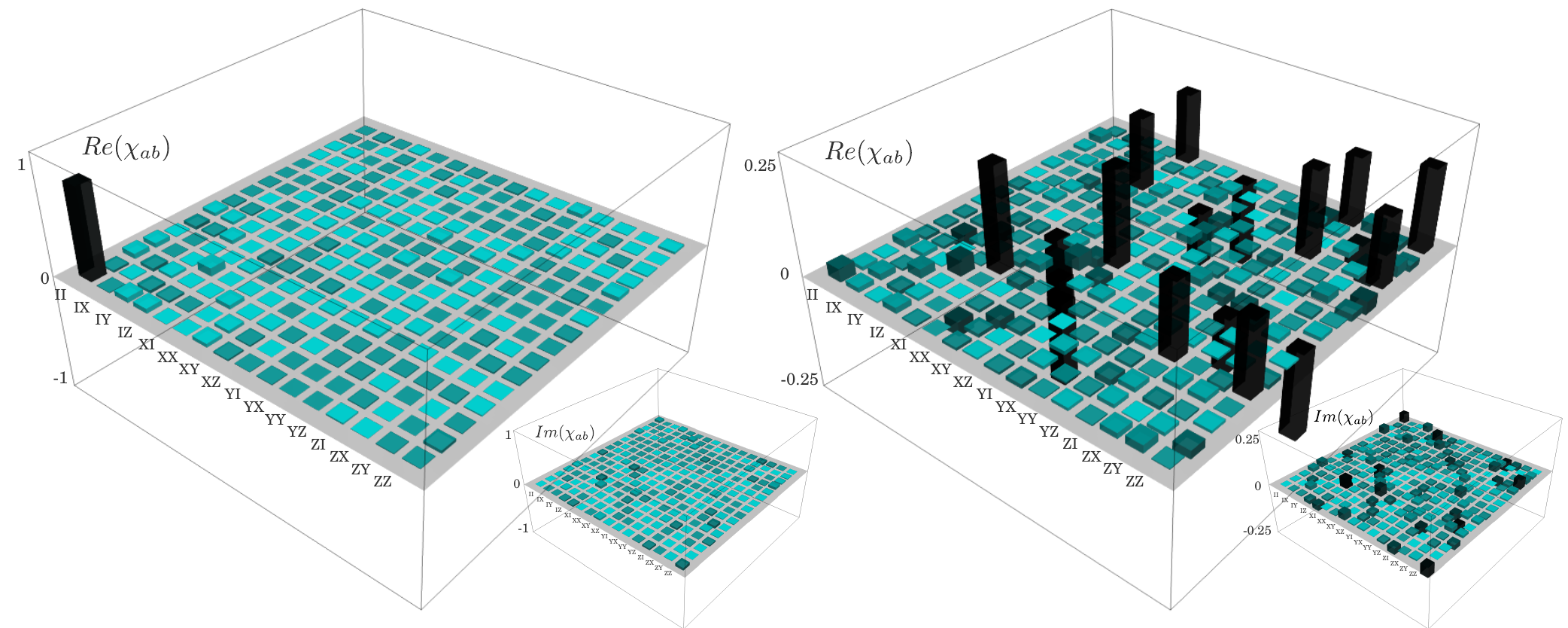, angle=0, width=0.8\textwidth}
\caption{\textbf{Full reconstruction of two channels}. The $\chi$ matrices for a) the identity process and b) a controlled $U_c$ process (see text).  Shown are the real and imaginary parts of each one all in excellent agreement with the expected theoretical results. The fidelity between these reconstructed processes and the ones reconstructed by the standard inefficient Nielsen and Chuang method are $91\%$ and $93\%$ respectively.}
\label{fig:results}
\end{figure*}

We implemented this method in an experiment to fully characterize several quantum channels affecting $n=2$ qubits. As a 2-design we used the $D(D+1)=20$ eigenstates of $D+1=5$ mutually unbiased bases (MUBs). In particular we chose the three separable bases whose generators are $X$, $Y$ and $Z$ for each qubit and two entangled ones generated by the operators $\{X\otimes Y, Y\otimes Z\}$ and $\{Y\otimes X, Z\otimes Y\}$.

\paragraph*{Photonic implementation.}

Several methods for complete or partial quantum process characterization have been demonstrated in different experimental setups  \cite{liu2008direct, altepeter2003ancilla,emerson2007symmetrized, diamond_limblad, compressedandy}. Here we implement our QPT method, which can selectively determine any parameter characterizing a quantum process, using a heralded single photon source obtained by Type-II parametric down conversion at a BBO crystal with a $405nm$ diode laser producing twin photons at $810nm$. Single photons encode two qubits: one associated to the polarization degree of freedom and the other to the path. As described in Figure \ref{fig:scheme} the experiment is divided in three stages: i) state preparation, ii) evolution with the quantum channel to be tomographed and iii) transition probability measurement. Polarization qubits are controlled with several wave plates while path qubits are controlled by three interferometers. Controlled operations are done with wave plates at different angles in each path. By appropriately combining a self-stable Sagnac interferometer, phase plates and wave plates we can prepare any desired state and measure any of the states of the $2$--design as required \cite{englert2001universal}. Figure \ref{fig:scheme} and its inset show the circuit equivalence of each part. State preparation and measurement are connected by an actively stabilized Mach-Zehnder interferometer in which the process is embedded (see supplementary material).

Various processes were studied: the identity; a unitary on the polarization qubit (a wave plate on both paths); an operation $U_c$ in which a different unitary is applied to the polarization qubit depending on the path qubit ($U_c=(I-Z)\otimes Z/2+(I+Z)\otimes X/2$ is implemented with wave plates at different angles in each path) and a noisy version of $U_c$ (noise in the path qubit is added by sweeping the Mach-Zehnder's phase). Figure \ref{fig:results} shows the full reconstruction of the $\chi$ matrix for the identity and $U_c$. In all measured processes we obtained excellent agreement with $\chi$--matrix of the ideal process and with the one measured by the standard method \cite{NC}. We computed the fidelity between the $\chi$ matrix obtained with our method and the one measured using Nielsen and Chuang's method. We obtained that for all the implemented channels such fidelity is above $90\%$. For full QPT we independently measured the values of all $\chi_{ab}$ elements. To do so, for each element we prepared the $20$ states in the $2$--design and measured their survival (and non-survival) probabilities. A simple algorithm to prepare states $(E_a\pm E_b)\ket{\phi_j}$ was developed. Full characterization of the $256$ elements of the $\chi$ matrix involves $256\times 40=10240$ transition probabilities. Fortunately many of them coincide and the number of different transition probabilities is much lower. In our case, a full characterization of a two qubit process requires only $140$ different experimental settings, each giving $4$ probabilities, resulting in $560$ transition probabilities to be measured (see supplementary material).

\begin{figure}
\begin{center}
\epsfig{file=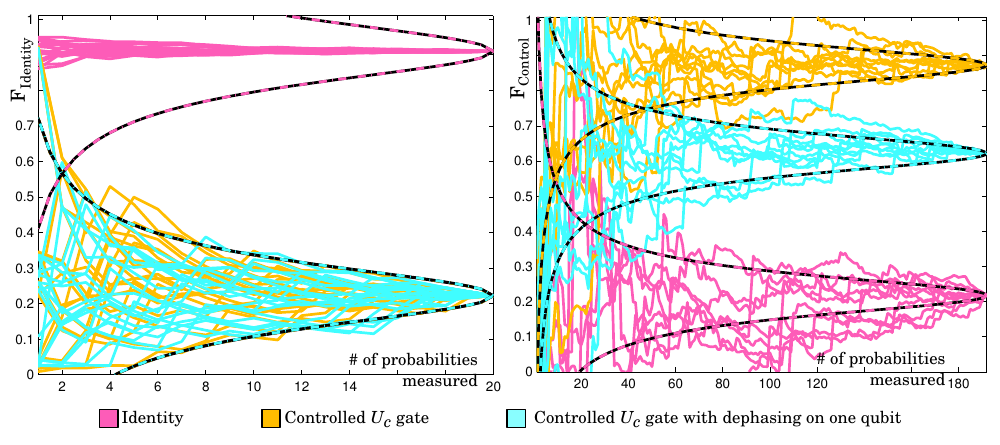, angle=0, width=0.45\textwidth}
\end{center}
\caption{\textbf{Efficient parameter estimation. }By determining only some parameters characterizing a quantum channel different relevant questions can be answered. These parameters can be determined exactly or approximated by sampling over only some of the possible measurements needed to fully determine them. Depending on the complexity of the question more or less measurements are needed. In any case these values can be determined statistically with an error scaling with the amount of measurements better than $1/\sqrt{M}$ eliminating the need for exponential amount of measurements. Shown are the fidelities of the measured processes to two different target processes: the identity (left) and a controlled gate $U_c$ (right) as a function of the amount of measurements done and for ten different choice of sampling order. Results for different channels show how quickly the estimation can differentiate between each one and converge to a value close to the exact one. Also shown in black and coloured dashed curves is the statistical maximum deviation expected for each result.
}
\label{fig:convergencia}
\end{figure}

However for the above full QPT, we do not take advantage of the most powerful aspect of this method: efficiency and selectivity. We also performed efficient  partial quantum process tomography measuring useful properties of the channel without fully determining the $\chi$ matrix. In this way we clearly show the advantage of our method over previous ones. Suppose we are interested in determining how close a given process is to a target process. A good measure of such distance is provided by the average fidelity of the channel obtained by composing the inverse target operation and the channel $\Lambda$. It is simple to show that this average fidelity is $F=(D Tr(\chi\tilde\chi) +1)/(D+1)$, where $\tilde\chi$ is the $\chi$--matrix of the inverse target channel. Such matrix can be simply obtained analytically and is typically small: For example, for the identity channel the only non-vanishing element is $\tilde\chi_{00}=1$. Therefore to estimate the fidelity of the identity we only need to estimate $\chi_{00}$. On the other hand, for the above mentioned controlled operation $U_c$, the $\tilde\chi$ matrix has $16$ non-vanishing elements ($4$ diagonal and $12$ off-diagonal ones). Therefore, to measure the fidelity of such channel we only need to estimate $16$ elements of the $\chi$--matrix. Moreover, such coefficients can be estimated with increasing precision by increasing the size of the sample. In Figure \ref{fig:convergencia} we show how these fidelities converge when the sample size is increased (curves correspond to different random choices for the order in which we sample over the $2$--design). In all cases we see that it is possible to decide if the channel $\Lambda$ is close enough to the target channel by making a number of measurements that is much smaller than the one required for full QPT. In such figure we also show that the same method reveals the presence of noise in the controlled operation. Not only an exact answer to such questions requires fewer resources than in previous methods \cite{NC, mohseni2007direct} but a good estimate can be obtained with a number of measurements that does not scale exponentially.

\paragraph*{Conclusion.}

The method presented accomplishes one of the fundamental tasks needed to build a general purpose quantum computer: it can efficiently and selectively characterize any quantum process.  The power of this quantum device  arises from the ability to prepare, sample and measure a special set of states. We showed that even for a two qubit system the strategy we presented displays clear advantages over preexisting methods. Our photonic experiment confirms this by identifying crucial information determining the nature of different quantum processes in a selective manner. The crucial task of estimating how close a certain evolution resembles a target process can be performed with our technique in a selective and efficient way. This, together with the estimation of the most important sources of errors affecting a quantum process are examples of the type of usage that this type of quantum tomographer will have.

%\bibliography{referencias}
%
%\bibliographystyle{Science}

%\begin{thebibliography}{10}

%\end{thebibliography}

Authors acknowledge J.Codnia and A.Hnilo for help and encouragement during different stages of this work. This work was funded by ANPCyT, UBA and CONICET.

\section*{Suplementary Material}

\paragraph*{Calculation of probabilities.}
For each configuration, clicks in each output were recorded during 10 seconds. Probabilities were calculated by normalizing this amount of clicks to a reference measured with the identity process and by preparing and measuring in the computational ($ZZ$) base. This reference was updated in periods of less than 25 minutes to account for slow fluctuations in the laser power.
All optical fibers on the signal side where single mode fibers. The average amount of maximum counts per second were: 1800000 on the herald, 2600 on the signal and 600 coincidence counts.

\paragraph*{Interferometer Stabilization.}
 The Mach-Zehnder interferometer was actively stabilized by a phase-sensitive closed-loop system. An auxiliary $405nm$ laser runs parallel to the $810nm$ single photon beam path and its interference signal is m
 easured with a photo-diode. Using this signal the position of one of the mirrors in the interferometer is actuated with a piezoelectric disc. The rotation of the $810nm$ wave plates only affects the $405nm$ reference signal in that its intensity varies considerably, but not the position of the maxima and minima. By making the piezoelectric disc oscillate with a small amplitude ($\approx 10nm$) at a frequency of $\approx 3kHz$ and comparing the relative phase of the measured oscillation with the forcing one, an analog circuit determines the necessary correction to the offset of the oscillation so as to maintain the interferometer always at a minimum of the interference fringes of the $405nm$ reference beam. With this simple phase sensitive technique we get stability of approximately $\lambda/20$ for the single photons.

\paragraph*{Imaginary parts of $\chi_{ab}$.}
In order to obtain the imaginary part of an off-diagonal element, we should consider a slight modification of the real part scheme shown above. First, we must consider the following two CP channels:

\beq
\tilde F_{ab}^\pm= \sum_j \langle\phi_j| \Lambda((E_a\pm i E_b)^\dagger |\phi_j\rangle\langle\phi_j| (E_a\pm i E_b)) |\phi_j\rangle.
\eeq
Then it is strightforward to obtain the imaginary part of one of the fidelities from equation 1 by measuring $\tilde F_{ab}^\pm$ and considering  $2Im(\tilde F_{ab})=\tilde F_{ab}^+- \tilde F_{ab}^-$.  
 
\paragraph*{Efficient state preparation.}
It is important to use a special $2$--design adapted to the basis of operators $E_a$. We choose $E_a$ as  generalized Pauli operators built as $n$--fold tensor product of the identity $I$ or one of the three Pauli operators ($X$, $Y$ or $Z$) on each qubit. These operators form the Pauli group that has $D^2$ elements (up to phases). This group can be partitioned into $D+1$ commuting subgroups each of which contains $D$ operators (including the identity) which are obtained as all possible products between $n$ independent generators. Each commuting subgroup defines an orthonormal basis of the Hilbert space, formed by the eigenstates of the operators in the set. These $(D+1)$ bases are mutually unbiased (MUBs)\cite{bandyopadhyay, lawrence-2002-65}: any state of one basis is an equally weighted superposition of all the $D$ states of any other basis. The set of all $D(D+1)$ states belonging to the $(D+1)$ bases form a $2$--design with special properties. This is the $2$--design we will use in our method, and we will denote it as $S=\{|\phi_i^{(\alpha)}\rangle, \alpha=0,...,D, i=1,...,D\}$. The index $\alpha$ labels the different MUBs and the index $i$ labels each state in each basis. For our method we will use the following properties of this $2$--design: a) Any state $|\phi_i^{(\alpha)}\rangle$ can be generated from any computational state (i.e. a joint eigenstate of all $Z_i$ operators) by an efficient quantum circuit\cite{BPP09}. b) Any Pauli $E_a$ is such that   $E_a|\phi_i^{(\alpha)}\rangle=|\phi_{i'}^{(\alpha)}\rangle$, that is to say, Paulis are translations within each basis. The transition rule, i.e., the expression that determines $i'$ as a function of $(i,\alpha, a)$ can also be efficiently obtained. In fact, such expression depends only on the commutation (or anti-commutation) relations between $E_a$ and the $n$ operators that are chosen as generators of the basis $\alpha$. c) Also the normalized state $|\Psi_{\pm,a, b, i}^{(\alpha)}\rangle=K(E_a\pm E_b) |\phi_i^{(\alpha)}\rangle$ can be generated efficiently from any computational state. The simplest way to do that is to prepare first a superposition of appropriately chosen computational states and later apply a change of basis. These tasks can be efficiently performed. The normalization constant $K$ is also efficiently computable.  \\
We need to prepare states of the form $(E_a+e^{i\beta}E_b)|\phi^{(\alpha)}_i\rangle$, where $\beta$ is a multiple of $\pi/2$ (odd multiples  are required for the measurement of the imaginary part of $\chi_{ab}$) and $|\phi^{(\alpha)}_i\rangle$ is one of the states from the $2$-design. To do this, we first fix an ordering of the states within each basis. On the computational basis ($\alpha=0$), we choose the lexicographic ordering. For any other basis we will use the convention $\ket{\phi^{\alpha}_i}=V^\alpha_0\ket{\phi^{\alpha}_i}$, where $V^\alpha_0$ 
is the corresponding change of basis operator from \cite{BPP09}.
The states we  prepare are then of the form
\begin{equation}
 \left(E_a+e^{i\beta}E_b\right)V^\alpha_0 X^{(i)}\ket{\phi^{(0)}_0}
\end{equation}
where $\ket{\phi^{(0)}_0}$ is the vector of the computational basis that has all zeros and $X^{(i)}$ is an operator that has $X$ on each qubit where the binary decomposition of $i$ has a one.

Since $V^\alpha_0$ is a Clifford group operator built with $O(n^2)$ Hadamard, CNOT and phase gates, it is efficient to compute how $E_a$ and $E_b$ transform into $\tilde E_a$ and $\tilde E_b$ under conjugation via $V^\alpha_0$\cite{BPP08}. This yields:

\begin{equation}
 V^\alpha_0 \left(\tilde{E_a}X^{(i)}+e^{i\beta}\tilde{E_b}X^{(i)}\right) \ket{\phi^{(0)}_0}
\end{equation}

And since the application of a Pauli operator on a computational basis state yields another state from that basis, the required state can be restated as 
\begin{equation}\label{eqFinPreparacionEstados}
V^\alpha_0 (\ket{\phi^{0}_m}+e^{i\gamma}\ket{\phi^{0}_n}) 
\end{equation}
which is the change of basis circuit acting on a state that is efficiently prepared via a Hadamard gate, $O(n)$ CNOT gates and at most three phase gates. The normalization constant is readily obtained from (\ref{eqFinPreparacionEstados}) as the norm of the state prior the application of the change of basis.

\paragraph*{Only 560 probabilities.} 
It is no surprise that full process tomography requires an exponential number of probability measurements. Just for diagonal tomography, each single coefficient is an average of 20 probabilities. Since there are 16 such coefficients this would require on the order of 320 probability measurements for full diagonal tomography. However, many of these probabilities are repeated. For instance, if we were to measure the diagonal $\chi$ coefficients corresponding to the operators $X\otimes X$ and $Y\otimes Y$, it is straightforward to see that both operators acting on the state $\ket{\phi^{(0)}_0}$ yield the same state, up to a phase. When it comes to off-diagonal tomography, many more of those probabilities are repeated. 
Since the $\chi$ matrix is defined by $O(D^4)$ real numbers and each each of those requires $O(D^2)$ probabilities to be obtained, it can be seen that full process tomography will require between $O(D^4)$ and $O(D^6)$ probability measurements, both exponential on the number of qubits. In the 2 qubits case it was found out that measuring only 560 probabilities was enough for full process tomography.

\end{document}